\documentclass[conference]{IEEEtran}
\IEEEoverridecommandlockouts

\usepackage[utf8]{inputenc}  
\usepackage{array}           
\usepackage{graphicx}        

\usepackage{dirtytalk}       
\usepackage{fontawesome5}    

\usepackage[T1]{fontenc}
\usepackage[nocompress]{cite} 
\usepackage{microtype}
\usepackage{balance}
\usepackage{graphicx}
\usepackage{enumitem}
\usepackage{amssymb}

\usepackage{url}
\usepackage{xcolor}
\definecolor{dark-red}{rgb}{0.5,0.1,0.1}
\definecolor{dark-blue}{rgb}{0.1,0.1,0.8}
\usepackage[colorlinks,linkcolor=dark-red,citecolor=dark-blue,urlcolor=dark-blue]{hyperref}
\usepackage[nameinlink]{cleveref}

\newcommand{\ttxplatform}{INJECT Exercise Platform}
\newcommand{\ttxplatformshort}{IXP}

\newcommand{\numteamstotal}{24}        
\newcommand{\numteamsanalyzed}{23}     
\newcommand{\numstudentstotal}{81}     
\newcommand{\numstudentsanalyzed}{76}  

\usepackage{tikz} 
\newcommand\copyrighttext{%
  \scriptsize \textcopyright 2026 IEEE. Personal use of this material is permitted. Permission from IEEE must be obtained for all other uses, in any current or future media, including reprinting/republishing this material for advertising or promotional purposes, creating new collective works, for resale or redistribution to servers or lists, or reuse of any copyrighted component of this work in other works. Cite this article as follows: V. Švábenský, J. Vykopal, S. Leelaluk, P. Čeleda, F. Okubo, A. Shimada. \textit{Assessment in Team Problem-Solving Exercises in Computing Education}. In Proceedings of the 56th IEEE Frontiers in Education Conference (FIE '26). Paphos, Cyprus, 2026. DOI: \textit{TODO add after the proceedings publication}.}
\newcommand\copyrightnotice{%
\begin{tikzpicture}[remember picture,overlay]
\node[anchor=south,yshift=5pt] at (current page.south) {\fbox{\parbox{\dimexpr\textwidth-\fboxsep-\fboxrule\relax}{\copyrighttext}}};
\end{tikzpicture}%
}

\begin{document}

\title{Assessment in Team Problem-Solving Exercises in~Computing Education
\copyrightnotice
\thanks{This research was supported by the Open Calls for Security Research 2023--2029 (OPSEC) program granted by the Ministry of the Interior of the Czech Republic under No. VK01030007 -- \textit{Intelligent Tools for Planning, Conducting, and Evaluating Tabletop Exercises}.}
}

\author{
    \IEEEauthorblockN{\textbf{Valdemar Švábenský}}
    \IEEEauthorblockA{Faculty of Informatics\\
    Masaryk University\\
    Brno, Czech Republic\\
    \texttt{valdemar@mail.muni.cz}\\
    \texttt{0000-0001-8546-280X}}
\and
    \IEEEauthorblockN{\textbf{Jan Vykopal}}
    \IEEEauthorblockA{Faculty of Informatics\\
    Masaryk University\\
    Brno, Czech Republic\\
    \texttt{vykopal@fi.muni.cz}\\
    \texttt{0000-0002-3425-0951}}
\and
    \IEEEauthorblockN{\textbf{Sukrit Leelaluk}}
    \IEEEauthorblockA{Center for ICT Infrastructure\\
    Yamaguchi University\\
    Yamaguchi, Japan\\
    \texttt{s\_leelaluk@yamaguchi-u.ac.jp}\\
    \texttt{0009-0003-3493-6878}}
\and
    \IEEEauthorblockN{\textbf{Pavel Čeleda}}
    \IEEEauthorblockA{Faculty of Informatics\\
    Masaryk University\\
    Brno, Czech Republic\\
    \texttt{celeda@fi.muni.cz}\\
    \texttt{0000-0002-3338-2856}}
\and
    \IEEEauthorblockN{\textbf{Fumiya Okubo}}
    \IEEEauthorblockA{Graduate School of Information Science and\\
    Electrical Engineering, Kyushu University\\
    Fukuoka, Japan\\
    \texttt{fokubo@ait.kyushu-u.ac.jp}\\
    \texttt{0000-0002-0077-9072}}
\and
    \IEEEauthorblockN{\textbf{Atsushi Shimada}}
    \IEEEauthorblockA{Graduate School of Information Science and\\
    Electrical Engineering, Kyushu University\\
    Fukuoka, Japan\\
    \texttt{atsushi@ait.kyushu-u.ac.jp}\\
    \texttt{0000-0002-3635-9336}}
}

\maketitle

\begin{abstract}
This full paper in the research-to-practice track presents methods for assessing student teams in tabletop exercises (TTXs). TTXs enable learner teams to prepare for workplace tasks and practice crisis responses, such as resolving cybersecurity incidents. While assessment is essential for determining how well teams achieve learning objectives, the complex, open-ended nature of TTXs often leads to delayed or incomplete feedback. TTX learning platforms can record teams' actions and communication; yet, leveraging these data to assess performance is underexplored. To address this gap, we compared two post-TTX team assessment methods—clustering and large language models (LLMs)—using an original dataset from 81 participants across two countries. We evaluated these methods against instructor-assigned scores based on standardized rubrics. Clustering grouped teams that approached TTX tasks similarly, enabling instructors to deliver faster, targeted feedback to teams within a cluster. This method was valid and reliable, with low computational requirements. LLMs used the standardized rubrics to assess teams' communication. While GPT-4o frequently disagreed with instructor scores, GPT-5.2 demonstrated considerably lower error. The researched methods have been integrated into INJECT, an open-source TTX learning platform, to support scalability and teaching practice. To encourage community adoption, we publicly share all datasets, software tools, and a full-fledged TTX scenario.
\end{abstract}


\section{Introduction}
\label{sec:intro}

Hands-on learning through \textit{tabletop exercises} (TTXs) is gaining traction in educational applications~\cite{fregeau2020use, wendelboe2020tabletop, husna2020does, SANDSTROM2014164}. In TTXs, which are grounded in simulation-based learning~\cite{chernikova2020simulation}, small teams of students collaboratively solve complex tasks within a limited time frame. All teams are seated in one room at separate tables, and instructors present shared tasks that each team addresses independently of other teams. Solutions are presented orally or submitted digitally, using tools ranging from simple online forms to specialized learning platforms.

TTXs are suitable for teaching \textit{cybersecurity}: a computing discipline~\cite{cc2023} that integrates technology, people, information, and processes to ensure operations in the face of adversaries~\cite{jtf-csec2017}. TTXs simulate cyber incidents: crisis scenarios that disrupt IT operations in an organization, such as a data breach. Team members collaborate to resolve the incident, tackling technical and non-technical challenges. As a result, TTXs foster competencies in realistic, workplace-like settings, preparing learners for the multifaceted demands of this rapidly evolving field. Due to these benefits, as well as the alignment with computing curricula~\cite{cc2023, Raj2022competencies}, TTXs have been applied in various cybersecurity training scenarios~\cite{Vykopal2024research, Angafor2024, hays2024usingllms, Muller2024, nakayama2024analyzing, Kavrestad2025}.

\subsection{Background, Current Gaps, and Problem Statement}
\label{subsec:intro-problem-statement}

In a TTX, each team's objective is to address incidents by completing specific tasks in accordance with appropriate procedures. In the customer data breach example, this involves actions such as blocking remote access to the database and contacting the system administrator. While each team faces the same incident scenario, their approaches can differ substantially in the activities they perform, their sequence, and timing. Therefore, to increase the TTXs' educational impact, it is essential to \textit{assess team performance} by evaluating which activities were completed and how well. Such an assessment can serve various educational purposes, including:
\begin{enumerate}
    \item grading teams in formal educational settings,
    \item supporting an immediate post-TTX debriefing (so-called \textit{hot wash}~\cite{Gafic2022exercises,us-hotwash}), and 
    \item providing feedback that highlights team strengths and areas for improvement, including how the teams deviated from the procedures expected by the instructor (which are not necessarily the only possible correct solution).
\end{enumerate}

However, assessing teams in TTXs poses many challenges. Teams may solve TTX tasks in various ways, complicating evaluation. Unlike established open-ended assessment tasks (e.g., essay grading~\cite{Xiao2025llmessay, Sessler2025llmessay}), TTX assessment lacks standardized frameworks. Lastly, team activity data vary in structure, format, and length, since teams submit written responses and supporting documents. In some TTXs, additional data include observer notes, audio/video recordings, or even IT system logs.

Traditionally, teams have been assessed manually through questionnaires and observations -- a practice still used today~\cite{Elvegard2024}. While manual data collection and analysis can yield valuable insights, it suffers from two major drawbacks.
First, TTX participants have to wait several days or even weeks to receive assessment-based feedback, which reduces its educational impact~\cite{Jeuring2022feedback}. By the time feedback is delivered, participants may have forgotten details of their actions or lack the time to attend a separate feedback session.
Furthermore, the time constraints force instructors to strike a difficult balance between speed and depth of assessment. Providing faster feedback often leads to superficial, generic assessments that fail to address individual teams' needs, while thorough feedback requires a lot of the instructor's time. This difficulty of delivering assessments that are both timely and meaningful has also been observed in other cybersecurity exercises~\cite{Mirkovic2020}.

We aim to improve the assessment by automatically analyzing TTX data using learning analytics methods. Using educational logs to provide insights into learning processes and outcomes has proven useful in various contexts~\cite{handbook-la2022,handbook-edm2010,Paiva2022automated}. However, these methods have not yet been validated in the TTX context. Our paper addresses this gap by empirically evaluating \textit{automated assessment of teams} and sharing practical experience from real-world TTXs. The proposed approach aims to speed up the assessment and deliver insights into team performance that are difficult to obtain manually. 

\subsection{Definition of Scope of Student Assessment in This Study}

We focus on assessment as characterized below, grounded in assessment theory and computing education frameworks~\cite{Raj2022competencies}.

\subsubsection{Teams, Not Individuals}
Our study examines \textit{overall team performance}, since a team is the fundamental unit. A team succeeds only when its members collectively resolve the tasks.
Analyzing individual contributions within a team, as in other non-TTX work~\cite{Feng2024}, is outside the scope of the current paper.

\subsubsection{Summative, Not Formative}
Team success is measured by completing tasks marked by predefined checkpoints called \textit{milestones} (e.g., \say{blocked the malicious website}).
Milestones serve as indicators of the learning objectives the team has met, and our assessment aims to summarize overall achievement.

\subsubsection{Exercise-Specific, Not Longitudinal}
The assessment relies only on \textit{data collected during a TTX}.
We do not incorporate historical participant data to ensure independence of external factors and support student data privacy.

\subsubsection{Post-Exercise, Not Intermediate}
During a TTX, instructors are busy facilitating and cannot divide their attention between intermediate assessments.
However, timely post-exercise feedback is crucial~\cite{Jeuring2022feedback,Mirkovic2020}. Assessment is thus conducted \textit{right after the TTX}, delivering instant results within minutes (instead of hours or days).

\subsubsection{Fully Automated, Not Manual}
To overcome the limitations of manual assessment (see \Cref{subsec:intro-problem-statement}), we utilize automated methods. Software tools process TTX data without human intervention, ensuring efficiency. Manual assessment is only used as a benchmark for evaluation in this study.

\subsection{Goals and Contributions of This Study}

Prior research on assessment in complex exercises (see \Cref{sec:related-work}) indicates that clustering and LLMs were effective in related educational contexts. Therefore, these methods may be viable for addressing the challenges in TTX assessment. However, the evaluation of the methods in the TTX context remains unexplored. To address this gap, the research aspect of our study is framed by two research questions (RQ):
\begin{itemize}
    \item[RQ1] \textit{[Quantitative] How does team assessment using (a) clustering and (b) LLMs align with instructor scores?}
    \item[RQ2] \textit{[Qualitative] Which insights for team assessment can be derived from TTX data using (a) clustering and (b) LLMs?}
\end{itemize}
Regarding the practice aspect of this research-to-practice paper, we bring various contributions and insights:
\begin{itemize}
    \item Based on our literature review (\Cref{sec:related-work}), this paper advances prior research and practice by evaluating automated TTX assessment using learning analytics methods.
    \item We conducted two different cybersecurity TTXs in authentic teaching environments (see \Cref{sec:exercise}), involving \numteamstotal\ teams of \numstudentstotal\ students in two countries (see \Cref{sec:methods}). These multi-national, multi-institutional study settings address many of the limitations of previous work~\cite{handbook-CER1}.
    \item We demonstrate and evaluate the assessment methods by analyzing student datasets. We compare the methods' alignment with manual scoring, their advantages and limitations, and their educational use cases (see \Cref{sec:results}).
    \item All materials and tools are available, enabling researchers and educators to adopt or adapt them (see \Cref{sec:conclusion}).
\end{itemize}

\section{Related Work in Team Assessment}
\label{sec:related-work}

Effective team-based problem-solving is critical in a wide range of domains~\cite{Taylor2024cps}, especially in high-stakes contexts such as healthcare~\cite{Zhao2024}, military~\cite{Pande2023}, and cybersecurity~\cite{nakayama2024analyzing, OConnor2024pwn, Narain2025practicalcybersecurity, Won2024Cybersecurity, Costa2025ctf, Gough2024remotecyber}. TTXs simulate crisis scenarios in these contexts, making team assessment vital for evaluating performance and learning outcomes. We examine prior research in this area, aligned with our RQs, to show how assessment challenges have been addressed in non-TTX contexts.

\subsection{Clustering in Team Assessment}
\label{subsec:related-work-clustering}

Clustering is an unsupervised machine learning technique widely applied in education~\cite{handbook-edm2010} to analyze team behavior, activities, and performance~\cite{dutt2017systematic}, as well as to provide educational feedback~\cite{Svabensky2022student}. 

For example, \cite{perera2008clustering} leveraged logs from software development group projects to enhance teamwork. They classified seven teams into three clusters based on activities such as commit frequency, ticket management, and wiki updates. The clusters' analysis uncovered diverse work strategies to inform teaching interventions.
In another programming context, \cite{sercce2011online} clustered written discussion data from 34 teams to analyze collaborative behaviors. The analysis revealed three clusters reflecting varying levels of collaboration, identifying teams with notably less interaction.
Similarly, \cite{jaiswal2021characterizing} examined team orientation in terms of goals, roles, and processes within a project-based learning environment. Clustering 23 teams showed that teams with balanced orientations achieved better academic performance compared to those with unbalanced orientations.

These studies demonstrate the utility of clustering when clear assessment objectives guide the evaluation. We advance prior work by extending this method to the open-ended TTX context, where such objectives are not explicitly defined. By clustering TTX data, we assess similarities and differences among teams. 

\subsection{Large Language Models in Team Assessment}
\label{subsec:related-work-llm}

The recent surge in the use of LLMs in education has introduced new benefits and challenges~\cite{kasneci2023chatgpt, Pankiewicz2025llm}. Among their many applications, the use of LLMs to analyze language data is relevant to our study. \cite{jia2024assessing} used GPT-3.5 to give feedback on student project reports. While the researchers cautioned about potentially unreliable outputs, model fine-tuning reduced hallucinations in the generated feedback. However, \cite{garg2024automated} evaluated GPT-3.5 and GPT-4 by coding data from interviews, and neither model achieved the expected reliability.

\cite{kakarla-borchers2025} used LLMs to assess open-text responses from tutors in educational scenarios. While this domain shares the ill-defined nature of TTX assessment, its objectives differ. Next, \cite{Snyder2024} assessed student programming collaboration by summarizing conversations of nine student pairs. \cite{Rudian2025llmfeedback} compared student perceptions of feedback generated by a teacher vs. an LLM. Lastly, \cite{FerreiraMello2025llm} provided evidence that traditional machine learning models outperformed LLMs in short answer grading.

Compared to prior work, we integrate rubrics from established cybersecurity curricular guidelines into the LLM-based assessment. The rubric aims to mitigate prior challenges and ensure that the TTX assessment aligns with learning objectives.

\section{Educational Context of Exercises in Our Study}
\label{sec:exercise}

This section outlines the properties of TTXs we used.

\subsection{Exercise Goals and Learning Objectives}
\label{subsec:exercise-goals}

From an educational-theory perspective, TTXs connect to the broader landscape of simulation-based learning~\cite{chernikova2020simulation}. Through simulations of authentic scenarios, students learn to address practical problems in a safe learning environment where failure has no negative consequences. Each team progresses with minimal to no guidance from the instructor, employing self-regulated learning (SRL) strategies. SRL is a well-studied framework~\cite{Li2024srl, Zhao2025srl, Cheng2025srl, Saint2024srl}, but not in the TTX context.

TTXs in our study simulate real-world cybersecurity incident handling, offering an immersive learning experience. The exercises mirror the tasks performed by a \textit{Computer Security Incident Response Team} (CSIRT) in medium to large organizations. Participants must prioritize and address multiple issues within a limited time frame (90 minutes), reflecting the pressures and dynamics of realistic cyber incident response.

The TTXs in this research have the same learning objectives, which align with the skills defined for the \textit{Incident Response} role in the well-established NICE Cybersecurity Workforce Framework~\cite{NICE}. In addition, the TTXs aim to develop dispositions~\cite{cc2023, Raj2022competencies} such as teamwork, decision-making, and professional communication -- both within the CSIRT and with stakeholders involved in the simulated incidents.

\subsection{Exercise Format and Content}

Participants assume the role of CSIRT members tasked with handling reports of incidents affecting the IT infrastructure of a simulated organization. \Cref{tab:ttxs} details the incident scenarios. The teams' responsibilities include: (1) discussing the reports to determine appropriate actions, (2) requesting information or collaboration from internal or external actors, and (3) using simulated tools for incident response. Although the TTX themes differ, they both address the overarching learning objectives defined in \Cref{subsec:exercise-goals}. Therefore, we study both TTXs jointly.

\begin{table}[t]
\caption{Similarities and differences between our exercises.}
\label{tab:ttxs}
\centering
\setlength{\tabcolsep}{4pt}
\footnotesize
\begin{tabular}{c|p{1.85cm}|p{1.45cm}|p{3.5cm}}
TTX & Cyber incident topic  & \# Milestones (important + secondary) & \# Tools (with representative examples in parentheses) \\ \hline
EXF      & Data \underline{exf}iltration by malware & 17 (5 + 12) & 11 (IP geolocation, IP blocking, log search, \dots) \\ 
PHI      & \underline{Phi}shing attacks campaign    & 18 (14 + 4) & 11 (DNS lookup, log search, email sender blocking, \dots) \\ 
\end{tabular}
\end{table}

The TTXs were conducted in-person, using the open-source \ttxplatform\ (\ttxplatformshort)~\cite{Svabensky2024from}. Prior to the TTX, participants completed a brief tutorial on using the platform and were assigned to teams (see \Cref{subsubsec:ttx-participants}). After a briefing, participants could ask questions and designate team roles, such as appointing a team member to interact with the \ttxplatformshort\ interface.

Once the TTX began, teams interacted solely with their teammates and the \ttxplatformshort, which delivered messages based on a predefined scenario. Instructors role-played as TTX actors who communicated with teams via email in the \ttxplatformshort. The TTX intentionally provided minimal guidance to simulate the uncertainty of a cyber incident. Participants received context (e.g., \say{Your organization is experiencing phishing attacks.}) and were thrown in the middle of the scenario. They had to triage written queries from stakeholders (e.g., \say{I opened this email attachment, and my computer stopped working. What should I do?}), evaluate the situation, prioritize actions, devise solutions, and write up relevant responses.

Each TTX included activities aligned with the learning objectives. These activities could sometimes be tackled in any order, reflecting the non-linear nature of real-world incident response. To further enhance authenticity, the \ttxplatformshort\ did not provide immediate feedback on teams' progress; this was discussed later in the \textit{debriefing}. After the TTX, teams received a scenario summary, and instructors supplemented it by explaining the suitable procedures. To provide timely, in-depth support to instructors and teams, our paper evaluates assessment methods that enhance the post-exercise phase.

\section{Research and Assessment Methods}
\label{sec:methods}

\begin{figure}[!ht]
    \centering
    \includegraphics[width=\columnwidth]{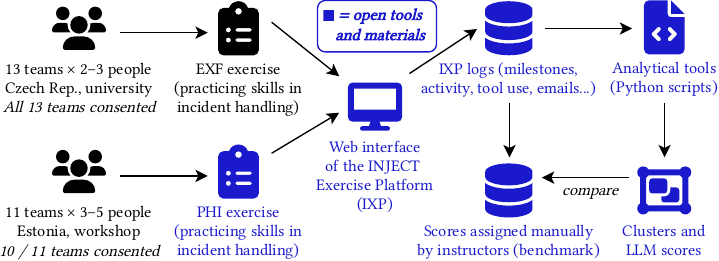} 
    \caption{Conceptual illustration of the research methods. Two distinct groups of teams independently completed the two TTXs in the same learning platform. Based on data collected from this platform, an assessment was conducted both manually and automatically, and the scores and outputs were compared.}
    \label{fig:method}
\end{figure}

\Cref{fig:method} provides an overview of our study, illustrating the important components and their interconnections. 

\subsection{Data Collection From Authentic Teaching Contexts}
\label{subsec:methods-data-collection}

We focus on computing students (including, but not limited to, those in dedicated cybersecurity programs) in higher and formal education. To support assessment generalizability, we gathered diverse data from various TTXs, students, and countries, as advised by computing education researchers~\cite{handbook-CER1}.

\subsubsection{Research Ethics}

We received a waiver from the ethics review board, i.e., the committee did not require explicit approval. All TTXs were conducted primarily for educational purposes. In addition, the instructor explained to the student teams that anonymized TTX activity data could be used for educational research. Each team indicated their consent as a whole via the \ttxplatformshort; declining consent had no negative consequences. This gave students the most control: if at least one team member was uncomfortable, the data were not stored.

\subsubsection{Participant Population}
\label{subsubsec:ttx-participants}

In the EXF exercise (see \Cref{tab:ttxs}), 36 cybersecurity students from a Czech university participated. The TTX was integrated at the end of a semester-long course on cyber incident handling, allowing students to apply the knowledge they had gained. The instructor divided the students into 13 teams to ensure balance based on their skill levels. While the course had 39 enrolled students, three absences resulted in 10 teams of three and 3 teams of two. 

PHI exercise was part of an event in Estonia to foster cybersecurity skills and collaboration between the two countries. Participants were 11 teams -- 5 teams of university students with a security background (three per team, self-selected), 5 teams of vocational school students (five seniors per team, selected by their teachers), and 1 team of five computing educators.

In total, \numstudentstotal\ participants in \numteamstotal\ teams completed the TTXs. However, one team declined consent for research, leaving \numstudentsanalyzed\ participants in \numteamsanalyzed\ teams in the dataset. Our group size aligns with or exceeds that of comparable state-of-the-art research; e.g., \cite{Feng2024} analyzed data from 15 teams of four, \cite{Bradford2023} used data from 10 teams of three, and \cite{Won2024Cybersecurity} had 4~teams of four to five. 
The TTXs were in English, which was a second language for all teams, though they were sufficiently proficient. Verbal communication within the team (out of scope of this study) was in the team's respective native languages. 

\subsubsection{Data Content}

The \ttxplatformshort\ logged each team's activity, e.g., milestone achievements, tool usage, and email responses. Communication between teams and instructors occurred only via the \ttxplatformshort. 
The data required filtering to use only \textit{teams’ activity}, ensuring the analysis was relevant to team assessment. We excluded data such as pre-scripted messages by the \ttxplatformshort, which were uniform across teams and would have caused noise.

\subsubsection{Instructor-Assigned Scores}
\label{subsubsec:manual-grading}

To establish a benchmark for automated assessment, two expert human instructors manually assessed the teams. This manual scoring is used solely for research evaluation in this paper to address RQ1. However, in a practical classroom setting, this manual scoring would not occur, which aligns with the objectives in \Cref{subsec:intro-problem-statement}.

The instructors used the \ttxplatformshort\ logs to assign numerical scores to each team. Two independent aspects of team performance were assessed -- milestone completion and communication. In both cases, the instructors followed pre-defined scoring rubrics (see below) to ensure consistency. Both rubrics assigned scores on a 3-step ordinal scale from 0 to 2 to ensure comparability.

First, a simple script developed by the instructor scored each \textit{team's achievement of TTX milestones} to obtain team scores. For each team and each milestone, we assigned a score of 2 for completing an important milestone, 1 for completing a secondary (optional) milestone, and 0 for missing a milestone. The final assessment for each team was an $m$-dimensional vector of these scores, where $m$ is the number of milestones in that TTX, as indicated in \Cref{tab:ttxs} (i.e., 17 or 18).

Second, both instructors assessed \textit{whether a team's written communication aligns with standards for cybersecurity work roles}. We based the assessment criteria on well-established cybersecurity curricular guidelines~\cite{jtf-csec2017}. These refer to a U.S. government-endorsed competency framework~\cite{us-writing-competency} with a rubric for assessing writing in professional contexts. For each of the rubric's three aspects (see \Cref{tab:results-llm} later), the instructors assigned a score of 2 if an aspect was satisfied, 1 if it was partially satisfied, and 0 if it was not satisfied. So, the assessment for each team was a 3-dimensional score vector.

\subsection{Automated Assessment of TTX Teams}
\label{subsec:methods-data-analysis}

To address our RQs, we evaluated automated methods. One author implemented them in Python (see code in \Cref{subsec:conclusion-materials}), another conducted a code review, and the next validated the results. For each method, we explain the use cases, algorithms, and input data. Then, we explain the method evaluation.

\subsubsection{Clustering}

In TTXs, where performance is reflected in the collaborative problem-solving process, behavior-based methods, such as clustering, are particularly well-suited for assessment.
Our goal was to assess \textit{how similar or different the teams' approaches were} in solving the TTX tasks. The resulting clusters help instructors identify teams that approached the tasks similarly, enabling them to provide assessment-based feedback to the whole cluster and to determine which approaches were more effective.
Since setting the suitable number of clusters in advance is difficult, we used density-based clustering with the DBSCAN algorithm. To reduce the impact of randomness caused by different seeds, we applied Unsupervised Consensus Clustering~\cite{Nguyen2007clustering}, averaging results from 50 iterations.
Features were based on \ttxplatformshort\ logs of which milestones teams reached (milestone IDs) and when (timestamps), including the sequence of actions and tool usage.

To quantitatively evaluate the outputs (RQ1), we examined team similarity within each cluster using manually assigned score vectors. We used the score vectors based on the milestones (not communication). Since the vectors have numeric components (0, 1,~2), we chose the Root Mean Squared Error (RMSE) because it captures both (a) score pattern closeness, indicating similar strengths and weaknesses, and (b) score magnitude, i.e., the performance on the important milestones matters more. We computed a pairwise RMSE between each member of a cluster (indicating the distance between two teams), then averaged the results for the whole cluster. This metric determines the validity of the automated output. As a comparative benchmark, our baseline was the average RMSE across all possible combinations of cluster groupings.

\subsubsection{LLM-based Approach}

An LLM mimicked the human assessment of a team's written communication. To guide the LLM-based assessment, we used the same rubric as in \Cref{subsubsec:manual-grading}, ensuring comparability with the human-based scores.
This rubric was integrated into a contextualized prompt that has been improved through multiple revisions. Then, we selected GPT-4o as the latest model at the time of our pilot study and GPT-5.2 for our validation study and provided the prompt via an API, tasking the model to assess how well the teams' communication adhered to the standards.
The input data consisted of the teams' incident response email messages. We tried several prompts, and the first few yielded results that differed from the expected format. Initially, the LLM produced a summary paragraph of high-level feedback for each team, rather than separately considering the three rubric criteria. After iteratively updating our strategy, the prompt that provided a desired output format is detailed below (shortened version to save space; see details in the code linked in \Cref{subsec:conclusion-materials}):

\begin{quote}
\textit{You are given email texts written by teams of learners during a tabletop exercise with the topic of cybersecurity incident handling. These teams practice their cybersecurity competencies in a professional context. For each team, evaluate their email communication with the stakeholders using the following three criteria, assessing whether the team:}
\begin{enumerate}
    \item \textit{Communicates information in a succinct and organized manner.}
    \item \textit{Produces written information appropriate for the intended audience.}
    \item \textit{Uses correct English grammar, punctuation, and spelling.}
\end{enumerate}

\textit{For each of the three criteria, output your assessment on the scale: Yes, No, Partially.}
\end{quote}

To quantitatively evaluate the LLM output (RQ1), we examined the error (i.e., the \say{disagreement}) between the instructor's and the LLM's ratings in the communication-based score vectors. We chose RMSE as the metric for the same reasons as for clustering, thereby also enabling comparison.

\section{Results and Their Discussion}
\label{sec:results}

\begin{figure}[!ht]
\centering
\includegraphics[width=\columnwidth]{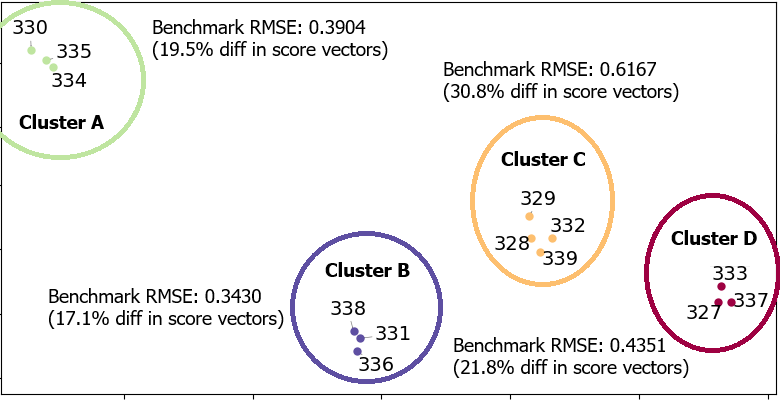}
\caption{Clustering in EXF. All 13 teams are identified by 3-digit IDs and grouped by activity similarity. The axes of the figure represent abstract dimensions in the clustering space, so their meaning is not directly interpretable.}
\label{fig:results-clustering}
\end{figure}

\Cref{fig:results-clustering} and \Cref{tab:results-llm} show example outputs of the automated assessment. \Cref{subsec:results-rq1} provides an overall statistical comparison of the results (RQ1). Then, \Cref{subsec:results-rq2} dives deeper into the specifics of each method (RQ2). \Cref{subsec:results-limitations} discusses limitations, and \Cref{subsec:results-implications} focuses on educational implications. Results are discussed separately for each exercise, highlighting differences across participant groups.

\newcommand{\yes}{{\small\faIcon{circle}}}
\newcommand{\partially}{{\small\faIcon{adjust}}}
\newcommand{\no}{{\small\faIcon[regular]{circle}}}

\begin{table*}[t]
\caption{Grading of communication of selected teams (out of all \numteamsanalyzed), demonstrating examples and differences between the final selected instructors' score and LLM's assessment. Grading scale: Yes (\yes\ = 2), Partially (\partially\ = 1), No (\no\ = 0). The last column also illustrates the effects on the RMSE metric based on the grade values in each row. This also provides context to \Cref{tab:results-rmse}.}
\label{tab:results-llm}
\centering
\setlength{\tabcolsep}{4pt}
\footnotesize
\begin{tabular}{c|p{8.6cm}|cc|cc|cc|c}
 & \textit{Note: The quotes in this column do not represent the entire grading decision. The conversations are much longer, and a broader context needs to be taken into account. The examples serve only illustrative purposes.} & \multicolumn{2}{p{1.6cm}|}{\textbf{Information succinct and organized}} & \multicolumn{2}{p{1.9cm}|}{\textbf{Appropriate for the inten\-ded audience}} & \multicolumn{2}{p{1.7cm}|}{\textbf{Correct grammar, spelling, punctuation}} & \\
\hline
Team & Example quote from the written communication & Human & GPT-4o & Human & GPT-4o & Human & GPT-4o & RMSE \\
\hline 
333 & \textit{The suspicious traffic was generated from user with compromised account} & \yes & \yes & \yes & \yes & \yes & \yes & 0.0000 \\ 
328 & \textit{We have serious reason to believe this is malware trying to spread.} & \partially & \yes & \yes & \yes & \yes & \yes & 0.5774 \\
329 & \textit{Please check if this scr ip is not in your competancy.} & \partially & \no & \partially & \partially & \partially & \no & 0.8165 \\
319 & \textit{!Check your email password and what you have sent!!!!!!!!!!!} & \no & \yes & \no & \yes & \no & \yes & 2.0000 \\
\end{tabular}
\end{table*}

\subsection{Comparison of Manual and Automated Assessment (RQ1)}
\label{subsec:results-rq1}

\Cref{tab:results-rmse} shows the evaluation results. For clustering, we assessed the cohesion of cluster membership according to the instructor's scores. For the LLM-based approach, we assessed its error compared to the instructor.

\begin{table}[!htb]
\caption{Differences between manual and automated assessment, expressed as average RMSE across teams per TTX and per method. The basis for measuring these values is defined in \Cref{subsec:methods-data-analysis}. Since the grading scale ranged from 0 to 2, the RMSE range is the same; a lower RMSE indicates lower error.}
\label{tab:results-rmse}
\centering
\scriptsize
\begin{tabular}{c|cc|c|c}
    & \multicolumn{2}{p{2.4cm}|}{\textbf{Clustering}: Assessing milestone completion} 
    & \multicolumn{1}{p{1.9cm}|}{\textbf{GPT-4o}: Assessing communication} 
    & \multicolumn{1}{p{1.9cm}}{\textbf{GPT-5.2}: Assessing communication} \\ \hline
TTX & Baseline    & RMSE        & RMSE & RMSE \\ \hline
EXF & 0.53 (27\%) & 0.45 (22\%) & 1.03 (51\%)  & 0.73 (37\%)  \\
PHI & 1.12 (56\%) & 0.79 (40\%) & 1.18 (59\%)  & 0.68 (34\%)  \\ \hline
Avg & 0.83 (41\%) & 0.62 (31\%) & 1.10 (55\%)  & 0.71 (35\%)  \\
\end{tabular}
\end{table}






\subsubsection{Clustering}

For $n$ teams, the number of possible non-empty cluster groupings is the $n$-th Bell number $B_n$. EXF had 13 teams, giving $B_{13}=27{,}644{,}437$ options. PHI had 10 teams, giving $B_{10}=115{,}975$ options. Based on \Cref{tab:results-rmse}, the baseline RMSE (average error across all these options) is much larger than for our clustering results. Therefore, our assessment has better internal cohesion of the clusters with respect to the manually assigned scores. We also note that this result applies across both TTXs. While the scenarios differ (leading to different baseline RMSEs), the clustering method is effective within each TTX. At the same time, some variance in the RMSE is essential. If the RMSE was 0, then all teams would be the same. So, to answer our RQ1(a), clustering sufficiently aligns with instructor scores and outperforms the baseline.

\subsubsection{LLM-based Approach}

We first compared the two instructors to establish the ground truth. In both exercises, the graders' RMSE was 0.32 (16\%), which represents differences in 10\% of assigned scores. In all cases, the differences were at most $\pm 1$ grade level. After one round of discussion, a final human score was established, favoring either instructor.

LLMs analyzed teams' email texts in natural language. However, since the GPT-4o's disagreement with the human-assigned grades was 55\%, the output essentially matched chance. Even though the LLM was guided by a clear prompt based on the standardized rubric criteria~\cite{us-writing-competency}, it often yielded surprising results (see, e.g., the last row in \Cref{tab:results-llm}). GPT-5.2 performed better, representing a substantial upgrade to 35\% error. Therefore, we answer RQ1(b): while a newer LLM performs better, the assessments are not sufficiently aligned with the instructor's judgment in the cyber TTX context.

\subsection{Qualitative Insights From Automated Assessment (RQ2)}
\label{subsec:results-rq2}

Next, we evaluated the qualitative utility of both methods. We link the obtained insights to educational use cases and show their practical benefits and limitations.

\subsubsection{Clustering}

Clustering grouped similar teams by compressing their data into two dimensions while preserving the relationships among features. Given the relatively small number of teams (13 and 10 in our data), it is expected to obtain few clusters. We observed from three to four clusters, each highlighting different patterns of team behavior and activity.

For example, the EXF data yielded four clusters based on teams' milestone achievements, see \Cref{fig:results-clustering}. Cluster~A included the highest-scoring teams that took a balanced approach, thoroughly addressing both technical and non-technical aspects. Cluster B also included high-scoring teams that excelled at procedural tasks but overlooked smaller technical interventions and secondary stakeholders. Cluster C was formed by the lowest-scoring teams. Although their performance was decent, they sometimes missed important milestones, did not use the correct tools, and lacked communication with TTX actors. Cluster~D contained average-scoring teams that used technical tools but forgot to contact some key stakeholders. This information can be used by instructors to enable cluster-differentiated feedback, for example, by recommending that teams in Cluster D contact actors such as the data protection officer, who is an important entity in the exercise context.

Another reason why clustering is valuable for instructors is that it provides a visual overview of teams that approached the tasks similarly. Moreover, outliers that do not belong to any cluster may indicate teams with unique approaches. For example, in PHI, only two teams were successful and formed a separate cluster. It was the only one to follow the entire incident response protocol, achieving even easy-to-miss milestones.

Next, our analysis revealed that the utility of time-based features, such as the sequence of tool usage, depends on the TTX content. In TTXs centered around a single incident (like in our study), these features provided relevant information for clustering. However, they may be less informative if the task completion order is more arbitrary.

Finally, we note a limitation of clustering. Features derived from email texts, such as common words and n-grams, were not useful for distinguishing between the teams. Many phrases -- mainly those related to cybersecurity processes -- were equivalent across most teams, resulting in a single cluster.

\subsubsection{LLM-based Approach}

Since the teams' written communication is in natural language, we assumed that an LLM would be well-suited for uniform assessment of teams' email content. However, GPT-4o's insights were not useful due to significant disagreement with the instructor; GPT-5.2 performed better. We attempted multiple prompts (see \Cref{subsec:methods-data-analysis}), but the first few produced unusable results. Throughout iterative improvements, we arrived at a prompt that appears to provide comprehensive coverage, and although the output looks reasonable at first glance, it often differs from the human expert's assessment.

We also experimented with categorizing the emails' emotional tone. However, the general LLM was biased to almost always interpret messages as exhibiting only anxiety-related emotions. Most likely, this stemmed from the prevalence of incident words (e.g., \say{cyber attack} or \say{data loss}) in the dataset. These results show that to obtain valid results, it is vital to frame the assessment within the TTX context or to use a domain-specific model, such as CyLLM~\cite{mai2025cyllm} for cybersecurity.

Another limitation of the assessment using public (non-local) LLMs is that it depended on querying an external service -- unlike clustering, which was executed locally. If the LLM became unavailable or its version changed, the assessment's validity and reliability would change as well.

\subsection{Limitations of This Study}
\label{subsec:results-limitations}

Regarding the teaching context, our study focused on students in cybersecurity or computing education programs who had at least a basic knowledge of cybersecurity. So, when using the same TTXs, our research findings may only reliably generalize to the same population type. Nevertheless, the TTX format may be used in many other teaching contexts (see \Cref{sec:intro}), potentially enabling the transfer of our methods to other classrooms and to entirely different curricula.

Regarding the research methods, one could argue that the sample size was relatively small; however, \Cref{subsubsec:ttx-participants} shows that our team size aligns with or exceeds those in comparable recent research in learning analytics. Nevertheless, a limitation in the methods is the restriction to specific LLMs. Since we do not have access to the LLM's internal workings, it is difficult to determine the reason the assessment sometimes differed. In a broader sense, this is a limitation of all black box methods.

\subsection{Implications for Educational Research and Practice} 
\label{subsec:results-implications}

In TTXs, student cohort sizes are constrained (tens of teams), but manual assessment remains slow and insufficient. Furthermore, it is complicated by the fact that teams can vary in size and come from diverse backgrounds, as evidenced by our dataset from real classrooms. Based on our study and teaching experience, we provide recommendations to inform educators and help improve TTX assessment practices.

\subsubsection{Design TTXs With Assessment in Mind}

For automated assessment to work, a thorough design of TTX learning objectives and activities is crucial. Otherwise, the log data may be less accurate in capturing the learning process. TTX creators should consider assessment goals during the design phase to enable higher-quality analytics in the post-TTX phase.

\subsubsection{Clustering Makes Assessment Feasible}

\Cref{subsec:results-rq1} showed that there are millions of options for team similarity in TTXs, making it impossible for the instructor to discover relevant groupings manually. Clustering provides a quick indicator of team similarity based on activity logs. For each TTX, all analytics were computed in under 2 minutes on a standard laptop, addressing the assessment challenges outlined in \Cref{subsec:intro-problem-statement}. Subsequently, the instructor can provide feedback to the whole cluster more quickly, saving time when focusing on the personalized specifics of each team.

\subsubsection{Be Cautious About LLM-based Assessment in TTXs}

Although professional communication is an important skill in TTXs, a general-purpose LLM sometimes failed to recognize nuances in this domain-specific context, even when guided by a standardized rubric in the prompt.

\subsubsection{Consider Ethical Implications of LLM-based Assessment}

As with all automated decisions, the output must be reviewed by a human instructor to ensure accountability and fairness. Our clustering considers only teams' activity logs, not any external factors, unlike a (potentially biased) black box like an LLM. Moreover, since clustering runs locally, learners' data are not shared with an external service, preserving privacy.

\subsubsection{Generalize the Principles to Other Team Learning Scenarios}

Our assessment methods are applicable not only in TTXs but can also be adapted to related computing education contexts. These include Capture the Flag (CTF)~\cite{Nelson2024dojo, Costa2025ctf}, Cyber Defense Exercise (CDX), and other complex collaborative problem-solving exercises in cybersecurity, e.g., in cyber ranges~\cite{Narain2025practicalcybersecurity, Won2024Cybersecurity, Gough2024remotecyber} and beyond~\cite{Zamecnik2023cps}. Although neither of these contexts is entirely similar to TTX, they are related in that they provide limited information to the student and may involve open-ended solutions, complicating assessment.

\section{Conclusions and Future Work}
\label{sec:conclusion}

This study evaluated the automated assessment of student teams who practiced cybersecurity incident resolution. While clustering enhanced assessment, LLMs were less effective in this task. Even a relatively recent model struggled to assess the teams' texts using a structured prompt, despite our revisions and inclusion of task context. Moreover, the LLM's opaque internal mechanisms limited our ability to diagnose the root cause. These findings highlight that, despite growing enthusiasm around LLMs, their applicability may be limited in certain domain-specific scenarios. Should LLMs be used for assessment, their outputs need to be carefully reviewed. 

Next, this study contributed the clustering framework to advance the state of the art of assessment in TTXs. We evaluated our methods on a unique dataset comprising two exercises and teams from two countries, thereby supporting the generalizability of the findings. Even though the scale of TTXs may seem relatively small (tens of teams), the instructor workload using this teaching method is high, limiting manual assessment. Our study identified and demonstrated this gap in the TTX literature and provided classroom-based evidence of clustering's applicability in supporting team assessment.

\subsection{Open Research Challenges}

Future research can investigate combining the two methods, potentially including other assessment methods, into a single ensemble framework. Here, aggregation (e.g., by voting) can enhance generalizability. Next, future work can focus on collecting audio or video data from TTXs to support assessment. Lastly, future work can develop cybersecurity-tailored LLMs~\cite{mai2025cyllm} to mitigate the limitations of general LLMs and focus on custom TTX assessment.

\subsection{Practical Tools and Resources for the Community}
\label{subsec:conclusion-materials}

To support the adoption of TTXs, we release numerous materials under a free, public, open-source license:
\begin{itemize}
    \item The \textit{PHI exercise definition} and others~\cite{ixp-scenarios}, which can be deployed in the free \ttxplatform~\cite{Svabensky2024from}. 
    \item The research \textit{dataset} from both TTXs~\cite{materials_dataset}, usable to research student behavior and task strategies.
    \item Comprehensive \textit{results} and visualizations (including those omitted from this paper due to space limitations)~\cite{materials_code}. 
    \item The \textit{analytical toolset} (Python code) to process the dataset and derive the results in this paper~\cite{materials_code}. The \ttxplatformshort\ team has since incorporated these assessment methods into the latest version of the platform; see \url{https://inject.muni.cz/}.
\end{itemize}

\section*{Acknowledgment}
This research was supported by the Open Calls for Security Research 2023--2029 (OPSEC) program granted by the Ministry of the Interior of the Czech Republic under No. VK01030007 -- \textit{Intelligent Tools for Planning, Conducting, and Evaluating Tabletop Exercises}.
We also thank Tomáš Hájek for his help with establishing the ground truth for the LLM assessment.

\bibliographystyle{IEEEtran}
\bibliography{references.bib}

\end{document}